\documentstyle[12pt]{article}
\newcommand{\eeqn}[1]{\label{#1}\end{equation}}
\newcommand{\eean}[1]{\label{#1}\end{eqnarray}}
\def\npb#1#2#3{    {\it Nucl. Phys. }{\bf B #1} (19#2) #3}
\def\plb#1#2#3{    {\it Phys. Lett. }{\bf B #1} (19#2) #3}
\def\prd#1#2#3{    {\it Phys. Rev. }{\bf D #1} (19#2) #3}
\def\prep#1#2#3{   {\it Phys. Rep. }{\bf #1} (19#2) #3}

\def\ptp#1#2#3{    {\it Prog. Theor. Phys. }{\bf #1} (19#2) #3}

\def\zpc#1#2#3{    {\it Zeit. f\"ur Physik }{\bf C #1} (19#2) #3}

\def\ibid#1#2#3{   {\it ibid. }{\bf #1} (19#2) #3}
\def\beq{\begin{equation}}
\def\eeq{\end{equation}}
\def\bea{\begin{eqnarray}}
\def\eea{\end{eqnarray}}
\def\ba{\begin{array}}
\def\ea{\end{array}}

\def\eq#1{{eq. (\ref{#1})}}
\def\eqs#1#2{{eqs. (\ref{#1}--\ref{#2})}}

\let\vev\VEV

\def\Im#1{\mathop{\rm Im}\{#1\}}
\def\Re#1{\mathop{\rm Re}\{#1\}}
\def\etal{{\it et al.}}
\def\ie{{\it i.e. }}
\begin{document}

\begin{titlepage}
\vspace*{-1.5cm}
\begin{center}
\hfill SISSA 173/93/EP
%\\[-1ex] \hfill OSU-TA-2/92
%\\[-1ex] \hfill MPI-Ph/91-64
\\[1ex]  \hfill November, 1993

\vspace{4ex}
{\Large \bf On Soft Breaking and CP Phases  }

\vspace{-.5ex} {\Large \bf in the Supersymmetric Standard Model   }

\vspace{6ex} {\bf  Stefano Bertolini } $^{b)}$
 \ \  and\ \ {\bf  Francesco Vissani } $^{a,b)}$

{\it
\vspace{1ex} a) International School for Advanced Studies, SISSA
%\\[-1ex]
\\[-1ex] Via Beirut 4, I-34013 Trieste, Italy
}

{\it
\vspace{1ex} b) Istituto Nazionale di Fisica Nucleare,
\\[-1ex] Sez. di Trieste, c/o SISSA,
\\[-1ex] Via Beirut 4, I-34013 Trieste, Italy
}

%\vspace{2ex} {\bf  Francesco Vissani    }

\vspace{6ex}
{ ABSTRACT}
\end{center}
\begin{quotation}
We consider a class of N=1 supersymmetric extensions of the Standard Model
in which the soft breaking sector is CP conserving at the GUT scale.
We study the question of whether the presence of explicit CP violation
in the Yukawa sector of the theory induces through renormalization
effects CP violating phases in the soft terms, which could lead to
observable effects. A clear pattern appears in the structure of phases
in the soft sector. In particular, the inclusion of
intergenerational mixing induces large phases in the flavour mixing
entries of the trilinear soft breaking terms, whereas the diagonal entries
remain real. A mechanism is proposed for generating through chargino
exchange a contribution to the neutron electric dipole moment which
can be a few orders of magnitude larger that that of the Standard Model,
although still out of reach of experimental tests. We comment on the
possible relevance of these phases for baryogenesis at the weak scale
in minimal supersymetric scenarios, recently considered in the literature.

\end{quotation}
\end{titlepage}
\vfill\eject

\noindent{\large Introduction}

In this letter we reconsider the problem of CP violation in supersymmetric
extensions of the standard Glashow-Weinberg-Salam
model (SM) of the electroweak interactions.
In particular, we want to address the question of whether the presence
of a CP violating phase ($\delta_{KM}$) in the Yukawa sector of the
supersymmetric theory may induce,
through the running of the
relevant parameters from the GUT scale to the Fermi scale,
CP violation in the soft breaking sector.

In general, it is well known that additional phases may appear in
the minimal supersymmetric version of the SM (MSSM) due to the complexity
of the soft SUSY breaking couplings. On the other hand, it is also
well known that the presence of these extra phases
(collectively $\delta_{soft}$)
would induce a 1--loop contribution to the electric dipole moment (EDM)
of the neutron, which for squarks and gluino masses of $O(100\ GeV)$
is too large and requires $\delta_{soft} < 0.01$ \cite{edm-susy-old}.
{}From this point of view it may be natural demanding the absence
of these phases altogether by assuming that the
soft breaking sector, as derived from the flat limit
($M_{Planck}\to \infty$) of an underlying supergravity theory,
is CP conserving. However, the presence of an explicit CP violation
in the Yukawa sector of the ``effective'' theory at the GUT scale
may reintroduce some CP violation in the soft breaking sector at the
Fermi scale, thus posing again the question of its relevance for
present day CP odd observables.

As a matter of fact, the presence of a small explicit
CP violation $\delta_{soft} \simeq 10^{-5}-10^{-6}$ in the trilinear
component of the soft breaking potential has been advocated by the
authors of ref. \cite{Denisetal} in order to trigger the
baryon--antibaryon asymmetry in the early universe in a
finite temperature MSSM scenario.
It is worth stressing that the correct sign
of the baryon--antibaryon asymmetry depends crucially on the
sign of the explicit soft breaking phase. It is therefore interesting
to study whether the $\delta_{KM}$ induced phases previously mentioned
have the correct size and sign to support such a scenario.

Our main conclusion is that within the simplified approach of ref.
\cite{Denisetal}, where flavour mixing in the effective potential
is neglected, no appreciable phases are induced in the relevant
soft breaking parameters. It turns out, however, that due to the presence
of flavour mixing
large CP violating phases may be induced in the
off--diagonal components of the trilinear soft breaking couplings.
We study the implications of such a large phases for the
EDM of the neutron and find that they induce a contribution
consistent with the present upper bound, although it can be as much as
four orders of magnitude larger than the SM contribution to the elementary
EDM of the quarks. The relevance of these off-diagonal components
for the analysis of ref. \cite{Denisetal} deserves a separate
discussion.

\noindent{\large The Minimal Supersymmetric Extension of the Standard Model}

We are here interested in the class of N=1 supergravity based
minimal supersymmetric
extensions of the SM, in which the spontaneous breaking
of $SU(2)\times U(1)$ gauge symmetry is achieved
via radiative corrections (RMSSM) \cite{RMSSM}. These models, implemented
in a Grand Unified (GUT) scenario, exhibit the least number of free parameters
and are consequently the most predictive ones.

The SUSY-GUT theory is
obtained from the $M_{Planck}\rightarrow \infty$ limit of
spontaneously broken N=1 supergravity \cite{nillesnanopoulos};
the flat limit leaves a globally supersymmetric lagrangian, corresponding
to the supersymmetrization of the chosen GUT model (minimally an $SU(5)$ GUT)
which is explicitly broken by soft terms \cite{girardellogrisaru}.
After integration of the heavy GUT fields the soft terms can be casted
in the following form:
\beq
  -\sum_{ij} m^2_{ij} z_i^* z_j -(f_{\Gamma^A}(z)+B \mu h_1 h_2-
  \sum_{\alpha} \frac{M_\alpha}{2} \lambda_\alpha \lambda_\alpha + h.c.)
\eeqn{soft-terms}
where $z_i$ denote all scalar fields present in the theory.
The first term is a mass term common to all the scalars, whereas
$f_\Gamma$ is the part of the superpotential that extends the standard
Yukawa potential
\beq
\begin{array}{ccc}
f_\Gamma&=& { H}_1^0 { D}_L^t \Gamma_D { D}_L^c+
{ H}_1^0 { E}_L^t \Gamma_E { E}_L^c+
{ H}_2^0 { U}_L^t \Gamma_U { U}^c_L\\
&-&
({ H}_1^- { U}_L^t \Gamma_D { D}_L^c+
{ H}_1^- { \nu}_L^t \Gamma_E { E}_L^c+
{ H}_2^+ { D}_L^t \Gamma_U { U}^c_L)
\end{array}
\eeqn{superpotential}
where the three $\Gamma_x$ ($x=U,D,E$) are
$3\times 3$ matrices in flavour space,
and the upper index $t$ indicates trasposition in flavour space.
Notice that in eq. (\ref{soft-terms})
the dimensionless couplings $\Gamma$ have been replaced with
the massive parameters $\Gamma^A$, and the superfields with their
scalar counterparts.
The third term of \eq{soft-terms}
arises as the scalar counterpart of the term
\beq \mu { H}_1 { H}_2=\mu ({ H}_1^0 { H}_2^0-
{ H}_1^+ { H}_2^-)\eeqn{mu-term}
present in the superpotential,
while the last one is a mass term for the gauginos.

If one assumes  having a flat K\"alher metric,
the form of the soft
supersymmetry breaking terms turns out to be quite simple at the
scale of local supersymmetry breaking, that we  assume to be
the SUSY-GUT scale required by  gauge couplings unification,
namely $M_X\simeq 3\cdot 10^{16}\ GeV$.
In fact at that scale  we have:
\beq
\Gamma^A_x=A_{G}\cdot \Gamma_x\ \ \ \ \ \ \ \ \ \
x=D,\ E,\ U
\eeqn{A-init}
where $A_{G}$ is a massive universal coefficient (henceforth
the subindex $G$ denotes GUT scale quantities). In addition,
each scalar in the theory gets the same mass term
\beq
m^2_{G}\cdot z^*_i z_i
\eeqn{m0-init}
This form ensures the absence of
large flavour violating effects in neutral currents at the low energy scale.
We will also assume that the three  gaugino masses  are equal at this
unification scale:
\beq
M_\alpha=M_{G} \ \ \ \ \ \ \ \ \ \ \ \ \alpha=1,\ 2,\ 3
\eeqn{M-init}

Following a widely used notation we will distinguish the superpartner
fields using a tilde (for instance ${\tilde u}_L$ is the scalar partner of
the left up quark $u_L$, both belonging to the superfield $U_L$).

\noindent{\large  Complex Parameters in the RMSSM}

We have already mentioned that the 4 parameters
$A_{G} ,\ B_{G} ,\ M_{G} ,\ \mu_{G} $ can be a-priori
complex parameters. In fact, by a $R-$rotation, with $R-$charges $Q_R=1$
for lepton and quark superfields
and $Q_R=0$ for the vector and the Higgs superfields,
the gaugino mass parameter $M_G$ can be made real;
moreover, multiplying by a common phase the 2
Higgs superfields (with opposite hypercharge)
$B_G\cdot\mu_G$ becomes real as well.
We conclude that, in addition to the usual Kobayashi-Maskawa (KM)
phase \cite{km}, there are at most two tipically supersymmetric
phases that are physically relevant,
say
\beq
\arg(A_{G} )\ \ \ \ \ {\rm and}\ \ \ \ \  \arg(B_{G} )=-\arg(\mu_{G} ).
\eeqn{2-phases}
These two parameters have an important impact in the phenomenology of
CP violation.
They can induce an electric dipole moment
of the quarks and leptons at the 1-loop level through
diagrams of the type shown in fig. 1 (let us recall that in the
SM the elementary edm of quarks arises at the third loop \cite{shabalin}).
If the masses of the particles
running in the loop are $O(100)\ GeV$ the EDM
of the neutron is predicted to be
\beq
d_n^{SUSY}=O(10^{-23})\  \sin\delta_{soft}\ \ e\ cm
\eeqn{d-susy}
where $\delta_{soft}$ is a typical SUSY phase (say $\arg (A)$).
This prediction has to be
compared with that obtained
in the Standard Model
from the elementary EDM of the up and down quarks
(for a review see ref. \cite{shabalin}) :
\beq
d_n^{SM}(quarks)=O(10^{-34})\ e \ cm.
\eeqn{d-sm-elem}
As a matter of fact, it is likely that the neutron EDM is dominated
by long-distance (LD) effects \cite{edm-sm}, which lead to:
\beq
d_n^{SM}(LD)=O(10^{-32})\ e \ cm.
\eeqn{d-sm}
Present experiments give the upper bound \cite{edm-exp}:
\beq
d_n^{EXP} <  12\cdot 10^{-26}\ e \ cm
\eeqn{d-exp}
which, as already mentioned, excludes values of the supersymmetric
phases larger than $10^{-2}$ (or else
requires squark masses to be at the $TeV$ scale \cite{oshimo});
A measure of a non-zero value of the EDM of the neutron
in the next generation of experiments would certainly
be a signal of new physics, among which a supersymmetric
scenario with superpartners at the reach of the future hadron colliders.

\noindent{\large  Running of the Soft Breaking Parameters }

If one believes that squark and gluinos are just around the corner,
then a most conservative
and natural assumption for CP violation in the soft sector
is that the two phases in eq. (\ref{2-phases})
vanish identically
at the GUT scale, due to CP conservation
in the sector responsible for SUSY breaking.
However in this case, attention must be paid to the KM
phase, explicitly present in the model,
which may induce analogous phases in the soft breaking sector
through the renormalization
of the soft breaking parameters.

The complete set of 1-loop renormalization group
equation (RGE) for the SUSY model here
considered can be found for instance in ref. \cite{falck}
and in explicit matrix form in ref. \cite{BBMR}.
{}From inspection of the relevant RGE
one easily realizes that the gaugino masses do not change their
phase during the evolution, so that, within our hypothesis, they remain
real; an analogous conclusion holds for the parameter $\mu $,
which depends on the matrices $\Gamma$
only through the trace of the hermitian and non-negative combinations
\beq
\alpha_x\equiv {1\over 4 \pi}\Gamma_x\cdot \Gamma_x^\dagger\ \ \ \ \ \ \ \ \ \
x=D,\ E\ or\ U\ .
\eeqn{alpha}

%The mass matrices for left (or right) squarks (or leptons)
%are hermitian  matrices; their diagonal elements are real.
%Due to the initial diagonal
%condition (\ref{m0-init}), the off-diagonal elements
%are suppressed (we will come back on these parameters in the following).

In order to discuss the evolution of the parameters $B$ and $\Gamma^A_x$
we find convenient to define the matrices $A_x$ in the following way:
\beq
A_x\equiv \Gamma^A_x\cdot \Gamma_x^{-1}\ \ \ \ \ \ \ \ \ \
x=D,\ E,\ or\ U
\eeqn{A}
Defining ${\dot A}\equiv {dA/dt}$, with
\beq
t\equiv {1\over 4 \pi} \log({Q\over Q_0})
\eeqn{t}
we obtain the following RGE:
\beq
\begin{array}{ccl}
\dot {A}_E&=&2\ (3 \alpha_2 M_2 +3 \alpha_1 M_1) I\\
    &+&2\ {\rm Tr} (A_E \alpha_E+3 A_D \alpha_D) I\\
    &+&5\alpha_E A_E + A_E \alpha_E\\
\dot {A}_U&=&2\ ({16\over 3}\alpha_3 M_3+3 \alpha_2 M_2 +
      {13\over 9} \alpha_1 M_1) I\\
    &+&2\ {\rm Tr}(3 A_U \alpha_U) I\\
    &+&5\alpha_U A_U + A_U \alpha_U
       +\alpha_D A_U-A_U \alpha_D+2A_D \alpha_D\\
\dot {A}_D&=&2\ ({16\over 3}\alpha_3 M_3+3 \alpha_2 M_2
       +{7\over 9} \alpha_1 M_1) I\\
    &+&2\ {\rm Tr}(A_E \alpha_E+3 A_D \alpha_D) I\\
    &+&5\alpha_D A_D + A_D \alpha_D
    +\alpha_U A_D-A_D \alpha_U+2A_U \alpha_U\\
\dot {B}&=& 2\ (3 \alpha_2 M_2+\alpha_1 M_1)\\
    &+&2\ {\rm Tr}(A_E \alpha_E+3 A_D \alpha_D+3 A_U \alpha_U)
\end{array}
\eeqn{A-rge}
The $A_x$ are a-priori generic $3\times 3$ matrices, and the
form of the initial conditions is given in \eqs{A-init}{M-init}.

Notice that one can study independently the two cases
$ a)\ M_{G}=0,\ A_{G}\ne 0$ and
$ b)\ M_{G}\ne 0,\  A_{G}= 0$, and that
the GUT value for $B$, namely $B_{G},$ is always additive.

In order to analyze the evolution of the parameters
we resorted to a numerical study of this system of RGE, coupled to the
RGE for $a)$ the gauge coupling constants $\alpha_i,$ $i=1,2\ or\ 3$,
$b)$ the gaugino masses $M_\alpha$
and $c)$ the ``Yukawa couplings'' $\alpha_x,$ $x=U,D,E$.

The initial values of the matrices $\alpha_x$ at the $M_Z$ scale
are calculable once the value of $\tan\beta$,
the value of masses of the quarks and of the lepton and the
Kobajashi-Maskawa matrix are assigned.
In fact,
by writing the matrices $\Gamma_x$ in biunitary form,
\beq
\Gamma_x=L_x^t\gamma_x R_x
\eeqn{biunit}
where $\gamma_x$ are $diagonal$ non-negative matrices,
and making unitary
redefinitions of the quark and lepton superfields,
one finds that the parameters
in \eq{superpotential} can be chosen to be:
\beq
\left\{
\begin{array}{ccc}
\Gamma_E&=&\gamma_E\\
\Gamma_D&=&\gamma_D\\
\Gamma_U&=&K^t \gamma_U
\end{array}
\right.
\eeqn{gamma-init}
where $K$ is the $3\times3$ Kobayashi-Maskawa matrix
({\it down-diagonal basis}).

Denoting by $v_i$ the two vacuum expectation values of the Higgs fields,
{\it i.e.}
$\vev{h^0_i}   =v_i$ ($i=1,2$),
%($\tan\beta\equiv v_2/v_1$)
the matrices $\gamma_x$
are related to the (diagonal) mass matrices
for leptons and quarks as following
\beq
\left\{
\begin{array}{ccc}
\gamma_E&=&{M_E/ v_1}\\
\gamma_D&=&{M_D/ v_1}\\
\gamma_U&=&{M_U/ v_2}
\end{array}
\right.
\eeqn{gamma}
This basis is very useful when performing the RGE analysis
since the initial conditions are given in terms of masses
and KM mixings and it leaves
the $SU(2)_L$ symmetry explicit, having applied
the same unitary rotation to the $U$ and $D$ superfields.

To switch to the superfield basis in which the terms of
the superpotential
with the neutral Higgs superfields
are flavour-diagonal ({\it i.e. the quark mass eigenstate basis}),
we just have to redefine
the $U_L$ flavour multiplet of superfields as
\beq
U_L\rightarrow K^\dagger U_L\ .
\eeqn{change-fl-basis}
In this basis the mixing matrix $K$ appears
\begin{itemize}

\item[i)] in
the interactions involving the charged vector superfields

\item[ii)] in the
interactions of the charged Higgs superfields
in \eq{superpotential},

\item[iii)] and in the analogous terms for the charged Higgs fields
in \eq{soft-terms}.

\end{itemize}
Our numerical results will be shown in this basis.

In order to obtain the numerical solution of the system of RGE
we used the following set of electroweak input parameters:
\begin{itemize}

\item[1)] $\tan \beta=v_2/v_1$ in the range $1\div 40,$

\item[2)] the $\overline{MS}$ running quark masses at $m_Z$ (for a recent
          review see ref. \cite{arasonetal}); in particular, we tested
          for $m_{top}$ in the range $100\div 200\ GeV$.

\item[3)] the values of the three KM angles and phase consistent
          with the results of the analysis reported in ref. \cite{Buras}.

\end{itemize}
At the GUT scale we spanned values for the SUSY soft breaking parameters
which lead to a consistent SUSY mass spectrum at the electroweak scale.

We wrote the set of the relevant RGE
in matrix form and left to a program of symbolic
manipulation the task to expand them in vector form, suited for using
the variable-step NAG algorithm for the solution of
ordinary differential equations.
The results of our numerical analysis revealed that,
for any choice of the parameters,
the amount of the imaginary part of $B$ and of the diagonal elements in
$A_x$, induced at the $M_Z$ mass scale is
completely negligible ($\Im{A_{ii}}/\Re{A_{ii}}<10^{-20}$),
whereas the off-diagonal elements of $A_U$
and $A_D$ can have large phases (although the matrices $A_U$ and $A_D$
remain, up to 1 part per $10^{-3}-10^{-4}$, hermitian).

The following numerical example illustrates the tipical patterns
we found. Consider
for instance $\tan\beta=10$ and $m_t=130\ GeV$. Then
the values of the Yukawa matrices $\gamma_x$ at the $M_Z$ scale read
\beq
\begin{array}{ccl}
 \gamma_U&=&{\rm diag}(0.14\ 10^{-4},0.40\ 10^{-2},0.75)\\
 \gamma_D&=&{\rm diag}(0.29\ 10^{-3},0.55\ 10^{-2},0.17)\\
 \gamma_E&=&{\rm diag}(0.29\ 10^{-4},0.75\ 10^{-2},0.98\ 10^{-1})
\end{array}
\eeqn{gamma-init-num}
while for the KM angles and phase let us take
\beq
 (\theta_{12},\theta_{23},\theta_{13},\delta_{KM})=(0.221,0.043,0.005,0.86)
\eeqn{km-init}
where we tried to
maximize the CP violation effect from the KM matrix
through large values of
the mixing between the first and the third family and
$\delta_{KM}$ of order $1$,
according to the present experimental constraints (see for instance
ref. \cite{Buras}).

Finally, for
$A_{G}=1\ GeV$, $M_{G}=0$
(notice that in this case the RGE solutions for $A_x$ scale
as $A_{G}$), and $B_{G}$ arbitrary we obtain
\beq
\begin{array}{ccl}
A_U&=&
     \left(
      \begin{array}{ccc}
         0.78,&    (-1.3-i0.7) 10^{-6},& (-1.0-i1.7) 10^{-5}\\
         (-1.3+i0.7) 10^{-6},& 0.78,&    -1.7 \ 10^{-4}\\
         (-1.0+i1.7) 10^{-5},& -1.7 \ 10^{-4},&  0.56
      \end{array}
     \right)   \\
A_D&=&
     \left(
      \begin{array}{ccc}
       0.99,&    (2.2-i1.3) 10^{-5},& (-5.1+i3.1) 10^{-4}\\
      (2.2+i1.3) 10^{-5},& 0.99,&    (3.1+i0.1)  10^{-3}\\
      (-5.1-i3.1) 10^{-4},& (3.1-i0.1)  10^{-3},& 0.90
      \end{array}
     \right)   \\
A_E&=&{\rm diag}(0.99,0.99,0.98)\\
B&=&-0.12 + B_G
\end{array}
\eeqn{instance}
Increasing the top mass up to $200\ GeV$ affects little the elements
of $A_U$, while it modifies by factors $2-3$ the entries of $A_D$.
In addition,
due to the $1/\cos\beta$ ($1/\sin\beta$) dependence of the
down-quark (up-quark) Yukawa couplings,
the off-diagonal elements of $A_U$
increase roughly as $(\tan\beta)^2$, while those of $A_D$ remain roughly
constant.
The two features above are in fact related and may be qualitatively
understood by analizing the interplay between up an down Yukawa couplings
in the RGE for $A_U$ and $A_D$, (eqs. (\ref{A-rge})).

As mentioned before, the matrices $A_U$ and $A_D$ turn out to be
quasi-hermitian
(non-hermiticity appears at the 1 part per $10^{-3}-10^{-4}$ level
in the off diagonal elements and we have neglected it).
Notice also that the matrix $A_E$ remains in any case diagonal, because of
the absence of any source of lepton flavour violation.

A consequence of the ``hermiticity'' of the matrices $A_x$ is
the reality of B (up to a part in $10^{-20}$). In fact,
the evolution of B depends on the trace of the product
of two hermitian matrices ($\alpha_y\cdot A_x$), which is real.

The fact that the diagonal entries of $A_x$ are to an extremely good
approximation real bears important implications for the phenomenology
of SUSY induced CP violation, which we will shortly discuss.
This feature can be understood as follows.
Suppose decomposing the matrix $A_x$
in its hermitian and antihermitian parts
%\beq
$A_x=A_x'+A_x''$ respectively.
%\eeqn{h-hbar}
It is then evident that the boundary conditions for $A_x$
at the GUT scale, \eq{A-init}, imply
%\beq
$A_{x,G}''=0$,
%\eeqn{hbar-init}
so that the antihermitian part can be generated only through $[A'_x,\alpha_i]$;
but, due to the diagonality of $A_{x,G}'$ also the commutator
is zero at the GUT scale.
Moreover, to get a non-zero value of
$(A_x'')_{ii}$ from this commutator one readily verifies
that a certain amount of non-diagonality in $(A_x')_{ij},$ $i\ne j$,
is also needed, which is only induced radiatively.

\noindent{\large  The Neutron Electric Dipole Moment}

We shall now consider some consequences of the previous results.
Since both $\mu$ and $B$ remain real,
also the mass parameter $m_3^2=-B \mu$ responsible
for the mixing of the two scalar higgses
remains real at the end of the running.
Analogously
the gaugino mass matrices, \ie the chargino and the neutralino mass
matrices, are real; this fact implies that
only real orthogonal matrices are needed to reach the basis of mass
eigenstates.
Also the sleptons mass matrices are real (and diagonal), so that the
CP violating effects in the leptonic sector, such as the
electron or muon EDMs are zero at 1-loop level in the RMSSM as in the SM.

Let us now consider the role the
complex off-diagonal parameters in the matrices $A_x$
for the squark mass matrices, and begin our discussion
with the up-squark matrix.
The $6\times 6$ scalar quark mass matrix can be written in terms
of $3\times 3$ submatrices as follows:
\beq
M^2_{{\tilde u}}=\left(
\begin{array}{cc}
M^2_{{\tilde u},LL}&  M^2_{{\tilde u},LR}\\
M^2_{{\tilde u},RL}&  M^2_{{\tilde u},RR}
\end{array}
\right)
\eeqn{m-u-squarks}
The two $LL$ and $RR$ blocks are determined evolving the
scalar mass parameters in eq. (\ref{m0-init}) with the appropriate RGE;
the parameters $A_x$ enter only the $LR$ (and $RL$)
blocks
\beq
\begin{array}{ccl}
M^2_{{\tilde u},LR}&=&(\Gamma^A_U\ v_1+\mu\ v_2\ I)\cdot \gamma_U\\
                   &=&(A_U+{\mu \over \tan\beta}\ I)\cdot M_U
\end{array}
\eeqn{mLR}
The basis of squark mass eigenstates is reached using the unitary
rotation $S_{\tilde u}$, defined by:
\beq
 S_{\tilde u} M_{\tilde u}^2 S_{\tilde u}^\dagger ={\rm diag}(m^2_{\tilde u_k})
\eeqn{m-u-diag}
It is useful to split $S_{\tilde u}$ into two $6\times 3$ submatrices
\beq
S_{\tilde u}\equiv (S_{\tilde u,L},S_{\tilde u,R})
\eeqn{LR-splitting}
which relate the the scalar partners of the left and right-handed quarks
to the scalar mass eigenstates.

Notice that in the case of the down squarks,
since $M^2_{{\tilde d},LR}$ is proportional to $M_d$, the unitary matrix
$S_{\tilde d}$
is block diagonal up to $O(M_D/m_{SUSY})$; as we will
see this fact suppresses the up-quark contribution
to the electric dipole moment of the neutron.

Let us finally come to the neutron EDM.
Using a non-relativistic quark model,
the elemntary EDM of the quarks $d_u$ and $d_d$.
are related to the dipole of the neutron according to:
\beq
d_n={4\over 3} d_d-{1\over 3} d_u.
\eeqn{nqr-model}
Let us consider first the d quark.
The dipole can be computed from the imaginary part of the amplitude
$A^{dd\gamma}_{LR}$ for the
chirality-flipping $d_R \rightarrow d_L +\gamma$ process according to
\beq
d_d= - {\rm Im}(A^{dd\gamma}_{LR})
\eeqn{d-ALR}

In the SUSY model there are no 1-loop contributions to the neutron EDM
induced by $W^+$ or $H^+$ exchange. The only possibility left is to
examine typically SUSY diagrams with squarks running in the loop
(together with gluinos, charginos or neutralinos), and
resort to that component of the amplitude
in which the needed helicity flip is realized in the loop.
{}From a qualitative mass insertion analysis of the relevant diagrams
it is easy to convince oneself that a non vanishing (and possibly
complex) LR mass insertion in the squark line is needed.
In fact, when gluinos (or neutralinos) are considered,
the down-quark EDM turns out to be proportional at the leading order
to $\Im{(A_D)_{11}}$, which is zero in the scenario here considered.
One might circumvent this problem by invoking flavour changing (FC) effects
at the gluino-quark-squark vertex
so to involve complex off-diagonal entries in the soft insertion.
However these FC effects are of radiative origin, and for squark and
gluino masses heavyer than $100\ GeV$ turn out to give quite a
small contribution to the dipole amplitude (see ref. \cite{BBMR}).

We are therefore lead to consider chargino exchange, with up-squarks
running in the loop. Analogously to the gluino case if one considers
the ``diagonal'' exchange of $\tilde u_{u,c,t}$ no effect arises
due to the reality of the diagonal entries of the trilinear soft breaking
parameters.
On the other hand, there exist now also ``non-diagonal'' components
of the amplitude at the leading order (see fig. 1)
which are proportional to off-diagonal
elements of the $A_U$ matrix and to $complex$  combinations of the
relevant KM mixings. In particular we may expect that the dominant
contribution arises through the exchange of the ``higgsino''
component of the chargino field, via the quark-squark "flavour-chain"
\beq
d_R\rightarrow (\tilde{u}_L,\tilde{c}_L)\rightarrow
\tilde{t}_R\rightarrow d_L
\eeqn{u-chain}
since we can take advantage of the presence of the large top Yukawa coupling.
For large $\tan\beta$, then, also the presence of the down Yukawa
coupling becomes important as an enhancement factor.

\noindent
More concretely, using the interaction lagrangian (see for instance
the appendix of ref. \cite{BBMR})
\beq
\begin{array}{ccrl}
{\cal L}_{\chi \tilde{u} d}&=&\overline{\tilde {\chi}^-_a}\ \tilde{u}^\dagger\
  [&P_L(-V_{a1}^*S_{\tilde{u},L}g+V_{a2}^*S_{\tilde{u},R}\gamma_u)K\\
  &&+&P_R(U_{a2}^*S_{\tilde{u},L}K\gamma_d)]\ d\ \ +\ h.c.
\end{array}
\eeqn{lagra-chi}
where $U$ and $V$ are the $2\times 2$ orthogonal matrices responsible
for the diagonalization of the chargino ($\bar\chi^-$) mass matrix (see
for instance ref. \cite{gunionhaber}),
we find that the contribute to the d-quark EDM is can be written as
\beq
 d_d={1\over (4 \pi)^2}
  {\sum_{a=1}^{2}}
     {1\over m_{\tilde{\chi}_a^-} } V_{a2} U_{a2}
  {\sum_{k=1}^{6}}
     {\rm Im}\left[
       K^\dagger \gamma_u\left(S_{\tilde{u},R}^\dagger{\cal F}
       \left({m^2_{\tilde u_k}\over m^2_{\tilde \chi_a}}\right)
       S_{\tilde{u},L}\right) K \gamma_d
     \right]_{11}\ e\ cm
\eeqn{dipole}
The function ${\cal F}(x)$ is given by
\beq
  {\cal F}(x)={1\over 6 (1-x)^3} (5-12 x+7 x^2+2 x (2-3 x) \ln(x) )
\eeqn{f}

A formula analogous to \eq{dipole} holds for $d_u$. However, the appearance
of the down quark mass matrix in $M_{\tilde d,LR}^2$ (compare with \eq{mLR})
suppresses the EDM of the up-quark compared to that of the d-quark.

To numerically estimate this effect we considered the lightest chargino to
be close to the present experimental limit, say $m_{\bar\chi_1}
\simeq 50\ GeV$, lightest squarks masses of $O(100\ GeV)$ and maximize the
amplitude in the remnant parameters. We therefore find that the
SUSY contribution to the neutron EDM in this class of models
generated by the elementary EDM of the quarks can be as large as
\beq
d_n^{SUSY}=O(10^{-30})\left(\frac{\tan\beta}{10}\right)\ e\ cm
\eeqn{result}
where the linear dependence on $\tan\beta$ for large $\beta$
comes from the presence of the down quark Yukawa coupling
(the dependence on the top mass in the studied range is weaker).

This result has to be compared with the analogous SM contribution
of \eq{d-sm-elem}. In spite of the absence of phases in the diagonal
entries of the trilinear soft breaking terms, the smaller off-diagonal
terms may induce an elementary quark EDM four orders of magnitude
larger than that of the SM, and give a neutron EDM still two orders
of magnitude larger than that induced by long distance effects,
\eq{d-sm}. It is obvious, however, that this mechanism cannot explain
a measurement of the neutron EDM close to the present experimental bound.

For what concerns the possible implications of these results for
the analysis of ref. \cite{Denisetal}, on the generation of
baryon-antibaryon asymmetry at the weak scale in a finite temperature
MSSM scenario, no phases of $O(10^{-5}-10^{-6})$ are induced in the
diagonal (top) entries of the trilinear terms as suggested by the
authors. However, the aforementioned analysis just neglects
intergenerational mixing in the calculation of the effective potential.
The possibility that the large phases induced in the off-diagonal entries
of $A_x$ may actually become relevant and trigger the transition
to a baryon dominated universe remains open. Answering this question
requires a detailed analysis which is beyond the scope of the present letter.

%\vspace{1cm}
%\noindent
%{\bf Acknowledgements}

%.....................................

\vspace{1cm}
\noindent
{\bf Figure Captions}

\noindent
Figure 1.
The leading SUSY contribution to the elementary EDM of the quarks
in the class of models described in the paper
is shown in the interaction eigenstate basis.
The photon is attached in all possible ways.


\begin{thebibliography}{99}

\bibitem{edm-susy-old}
W. Buchm\"uller and D. Wyler, \plb{121}{83}{321};
F. de l'Aguila, M.B. Gavela, J.A. Grifols and A. Mendez, \plb{126}{83}{71};
J. Polchinski and M.B. Wise, \plb{125}{83}{393};
E. Franco and M. Mangano, \plb{135}{84}{445}.

\bibitem{Denisetal}
D. Comelli and M. Pietroni, Phys. Lett. B 306 (1993) 67;
D. Comelli, M.Pietroni and A. Riotto, preprint SISSA-93-50-A.

\bibitem{RMSSM}
K. Inoue, A. Kakuto, H. Komatsu and S. Takeshita,
\ptp{68}{82}{927}; \ibid{71}{84}{413};
L.E. Ibanez and G.G. Ross, \plb{110}{82}{214};
H.P. Nilles, \plb{118}{82}{193};
L. Alvarez-Gaum\'e, M. Claudson and M.B. Wise, \npb{207}{82}{96};
L. Alvarez-Gaum\'e, J. Polchinski and M.B. Wise,
\npb{221}{83}{495}.

\bibitem{nillesnanopoulos}
H. P. Nilles, \prep{110}{84}{1}; A.B. Lahanas and D.V. Nanopoulos,
\prep{145}{87}{1}.

\bibitem{girardellogrisaru}
L. Girardello, M. T. Grisaru, \npb{194}{82}{65}.

\bibitem{km}
M. Kobajashi and T. Maskawa, Prog. Theor. Phys. 49, (1973) 652.

\bibitem{shabalin}
E.P. Shabalin, Sov. Phys. Usp. 26 (1983) 297.

\bibitem{edm-sm}
D.V. Nanopoulos, A. Yildiz, and P.H. Cox, Phys. Lett. 87B (1979) 61;
I.B. Kriplovich and A.R. Zhitnisky, Phys. Lett. 109B (1981) 490;

\bibitem{edm-exp}
I.S.Altarev et al., JETP Lett. 44 (1986) 460;
K.F. Smith et al., Phys. Lett. B234 (1990) 2347.

\bibitem{oshimo}
Y. Kizukuri and N. Oshimo, Phys. Rev. D46 (1992) 3025.

\bibitem{falck}
N.K. Falck, \zpc{30}{86}{247}.

\bibitem{BBMR}
S. Bertolini, F. Borzumati, A. Masiero and G. Ridolfi, \npb{353}{91}{591}.


\bibitem{arasonetal}
H. Arason \etal, \prd{46}{92}{3945}.

\bibitem{Buras}
A.J. Buras and M.K. Harlander, in {\it Heavy Flavours}, Eds. A.J. Buras
and M. Lindner, World Scientific, Singapore, 1992.

\bibitem{gunionhaber}
H.E. Haber and G.L. Kane, Phys. Rep. 117 (1985) 75;
J.F. Gunion and H.E. Haber, Nucl. Phys. B272 (1986) 1.

\end{thebibliography}
\end{document}